# Evaluation of the blackbody radiation shift of an Yb optical lattice clock at KRISS

Myoung-Sun Heo, Huidong Kim, Dai-Hyuk Yu, Won-Kyu Lee, and Chang Yong Park*

Korea Research Institute of Standards and Science, Daejeon 34113, Republic of Korea

*Corresponding author e-mail: cypark@kriss.re.kr

**Abstract**

As optical clocks are improved to reach the frequency uncertainty below the $10^{-17}$ level, the frequency shift due to the blackbody radiation (BBR) has been one of the major systematic effects hindering further improvement. To evaluate the BBR shift of an Yb optical lattice clock at KRISS, we installed an in-vacuum BBR shield and made radiation thermometry using a black-coated-sphere thermal probe. After we quantitatively measured the conduction loss of the thermal probe and the effects of all the external radiation sources, we determined the temperature at the atom trap site with an uncertainty of 13 mK, which corresponds to an uncertainty of 0.22 mHz in the clock frequency (a fractional frequency of $4.2\times10^{-19}$). The total uncertainty of the BBR shift including the atomic response is $9.5\times10^{-19}$.

1. Introduction

Stark shifts due to the blackbody radiation (BBR) in most optical clocks are at the 1-Hz level [1]. This shift is one of the significant effects contributing to the frequency uncertainty and also affects the long-term stability of optical clocks. The BBR shift of the clock transition ($\Delta\nu_{\text{BBR}}$) can be expressed as [2]:

$$\Delta \nu_{\text{BBR}}(T_{\text{atom}}) = \Delta \nu_{\text{stat}} \left(\frac{T_{\text{atom}}}{T_0}\right)^4 + \Delta \nu_{\text{dyn}} \left[\left(\frac{T_{\text{atom}}}{T_0}\right)^6 + \mathcal{O}\left(\frac{T_{\text{atom}}}{T_0}\right)^8\right]. \tag{1}$$

where $T_{\text{atom}}$ is the temperature at the atom trap site, $\Delta \nu_{\text{stat}}$ is the coefficient related to the differential static polarizability between the clock states, $\Delta \nu_{\text{dyn}}$ is the coefficient associated with the small dynamic correction, and $T_0$=300 K. Since $\Delta \nu_{\text{stat}}$ is well known [2-4], the uncertainty in the BBR shift is mainly determined by the uncertainties of $T_{\text{atom}}$ and $\Delta \nu_{\text{dyn}}$. Much effort has been made for more accurate estimation of $T_{\text{atom}}$ during the last decade. The BBR shift uncertainty of 10[-17] to mid 10[-18] level was attained by measuring the temperature distribution around the vacuum system and by estimating the BBR from the heat sources [5−10]. To further reduce the BBR shift uncertainty to a low 10[-18] or 10[-19] level, well-controlled thermal environments were required. For example, the clock operation in cryogenic environments could reduce the BBR shift and its uncertainty [11−13]. Another method was to install an in-vacuum thermal shield at room temperature for an isothermal environment, and the additional inhomogeneous heat sources were dealt with a finite-element (FE) radiation analysis using effective solid angles [2,14,15]. Also, there has been a method of measuring $T_{\text{atom}}$ directly using a thermal probe [16] (and additional FE analysis in [17]) in a well-controlled thermal environment. In case of ion clocks [18, 19], the FE analysis with validation using a thermal imaging camera could reduce the uncertainty to 10[-18] level.

This paper describes the BBR shift evaluation of KRISS-Yb2, a newly developed Yb optical lattice clock at KRISS (Korea Research Institute of Standards and Science). We installed an in-vacuum thermal shield for homogeneous temperature distribution around the atoms. This thermal shield was similar to that used in [2] but was modified to have one side exposed in the air for more convenient temperature control. The temperature at the position of atoms was measured using a black-sphere thermal probe (BSTP). Using this thermal probe, we also obtained the effective solid angles of dominant heat sources experimentally. This approach enabled us to determine directly $T_\text{atom}$ without knowing exact values of geometric view factors and emissivities. The estimated uncertainty of $T_\text{atom}$ was 13 mK.

## 2. In-vacuum blackbody radiation shield

To evaluate the BBR shift accurately, we installed an in-vacuum BBR shield chamber made of copper inside a dodecagon titanium main chamber, as in figure 1. The outer rim of the BBR shield served as a gasket of the CF fitting on the the main vacuum chamber, as in figure 1(a). The BBR shield chamber was composed of the body part and the cap part, and they were assembled together with titanium bolts. Active temperature control of the BBR shield was carried out using a pair of heater films on the side wall of the body part (red lines in figure 1(a), which was exposed in the air as depicted in figure 1(a). Its temperature was controlled to be slightly higher than the room temperature. Two identical heater films with the same amount of currents in the opposite direction were overlapped to minimize the magnetic field induced by the currents. There were six platinum resistance thermometers (PRTs) installed in three deep holes and three shallow holes on the air-side of the BBR shield chamber to monitor the temperature. These PRTs and the precision resistance meter (milliK, Isotech) were calibrated with an uncertainty of 12.5 mK by the temperature standard group at KRISS. The PRT at the center acted as a reference probe for active temperature stabilization. The inner surface of the BBR shield was black-coated with Ultrablack® of Acktar [20] (emissivity > 0.98 in the wavelength range 3 ~ 10 μm) to reduce unwanted reflections. The QWP-mirror for retroreflection of cooling laser beams and the front window in figure 1(a) were ITO-coated and will act as electrodes to generate an electric field for the evaluation of the DC Stark shift which may occur due to the unwanted patch charges. As seen in figure 1(b), four additional electrodes were installed on the cap part of the BBR shield for electric fields along the other two orthogonal axes.

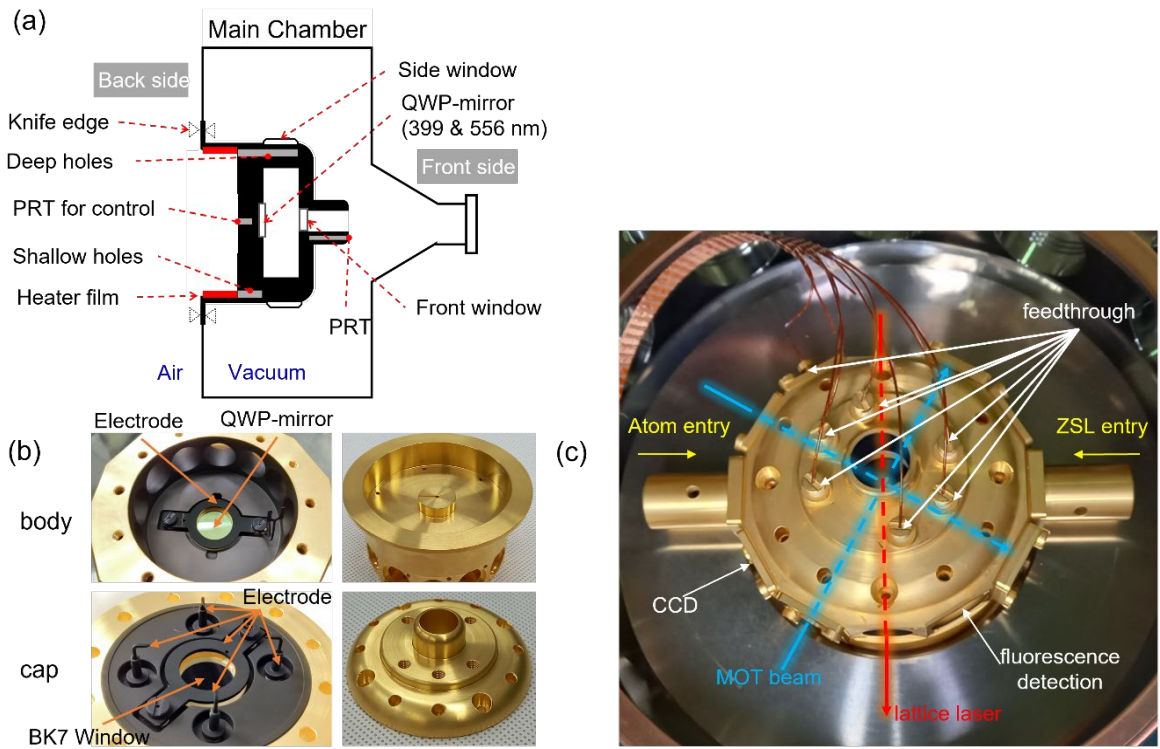

**Figure 1.** (a) The side-view schematic diagram of the BBR shield installed inside the main chamber. The BBR shield was knife-edge sealed to the main vacuum chamber. The PRTs were installed in the deep or shallow holes to monitor the temperature of the BBR shield. PRT: platinum resistance thermometer, QWP-mirror: mirror-coated quarter waveplate at 399 nm and 556 nm for the magneto-optical trap of Yb atoms. (b) Pictures of the BBR shield. It comprised the body and the cap. Upper left: inner view of the body part, upper right: outer view of the body part, lower left: inner view of the cap part, lower right: outer view of the cap part. The outer surfaces of the BBR shield were gold-coated to reduce the emissivity of the surface so as to minimize the effect of the external heat radiations and the temperature gradient of the chamber. The inner surfaces of the BBR shield were black-coated with Ultrablack® of Acktar to reduce unwanted reflections. Six electrodes were installed to evaluate the DC Stark shift

along three axes. Four electrode pins on the cap (lower left) were used to apply electric fields along two orthogonal axes. An electrode plate attached on the ITO-coated front window and another one attached on the ITO-coated QWP-mirror (upper left) were for the electric field along the remaining axis. (c) The picture of the BBR shield installed inside the main chamber. On the side of the body there are twelve apertures for optical accesses. See text for details. At the cap, there is one aperture covered with a BK7 window.

The in-vacuum temperature-stabilized BBR shield can ensure a homogeneous thermal environment. For the clock operation, we needed several optical and open accesses. The body part of the BBR shield has twelve 16-mm-diameter apertures. As seen in figure 1(c), two of them along the horizontal axis were open for the atomic beam and the Zeeman slowing laser, and they were extended by the 35-mm-long copper tubes with 8-mm inner diameter to reduce the BBR through the open apertures. Two apertures along the vertical axis were closed with copper plates with 4-mm-diameter holes at the center for the access of the lattice laser (the red dahed line in figure 1(c)). The blue dashed lines in figure 1(c) represent the magneto-optical trap (MOT) lasers, each of which transmits through a 4.5-mm-thick BK7 window and retroreflected on the QWP-mirror. Copper plates covered the rear sides of these two QWP-mirrors to block the external radiation. One aperture covered by a 4.5-mm-thick BK7 window was used for the viewport of CCD camera and another one covered by 10-mm-thick lens with the focal length of 35 mm for a fluorescence detection. The remaining two apertures were blocked by copper plates, one of which had an electrical feedthrough connected to the electrode on the QWP-mirror. The cap part of the BBR shield has one 16-mm-diameter aperture covered by a BK7 window. All the windows and a lens were made of BK7 glass to suppress the transmission of the external radiation above 3 μm, and their surfaces were coated with indium-tin-oxide (ITO) material to drain out electrical charges from their surfaces and generate the electric field.

In summary, the BBR shield chamber has four apertures open for the entrance of atoms and the lattice laser beam, and five apertures (four on the side and one on the top) closed with BK7 optics. These apertures on the BBR shield can introduce additional thermal radiation different from the temperature of the BBR shield. Also the low thermal conductivity of BK7 widnows induces the temperature inhomogenity of the BBR shield chamber. Those effects were investigated using the BSTP described in the following.

## 3. In-vacuum black-sphere thermal probe

Although the BBR shield drastically reduces the thermal environment's inhomogeneity, we cannot neglect the effects of external heat radiation sources and the residual inhomogeneity of the temperature and emissivity. To determine $T_{\text{atom}}$ exactly, one usually needs to know exact values given for geometric view factors and emissivity of all the external heat sources, which is sometimes not feasible. Therefore, we decided to directly measure the temperature at the atom trap site using an in-vacuum black-coated spherical thermal probe (BSTP) and estimated the effective solid angles of dominant heat sources. We will discuss the design considerations of the BSTP, such as geometry and emissivity. Then we will describe its systematic errors.

The BSTP is a black-coated (Ultrablack®, Acktar) 10-mm-diameter copper sphere with a PRT sensor (PT-111, Lakeshore) as in figure 2. The PRT sensor head (1.8 mm diameter and 5.0 mm length) was glued in the hole of the copper sphere using a thermally conductive epoxy. The resistance of the PRT was measured using four 200-mm-long 0.1-mm-diameter enameled Manganin® leads connected to the electrical feedthroughs on the top of the main chamber using the 4-point probe method. The BSTP was hung by gravity and was located at the center of the BBR shield. These Manganin wires are electrically conductive and thermally insulating with low thermal conductivity of 22 W/(m·K) at 300 K, which is about half the value of a widely used phosphor bronze wire.

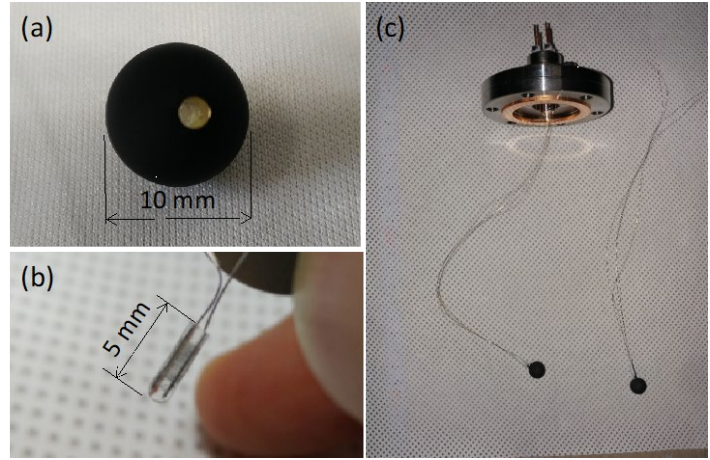

**Figure 2.** (a) The In-vacuum black spherical thermal probe (BSTP). The BSTP was made of copper and has a hole for a PRT sensor head. The surface of the BSTP was black-coated with highly emissive material (Ultrablack®, Acktar), (b) Picture of the installed PRT sensor (PT-111), (c) The BSTP was connected to the electrical feedthrough via four Manganin wires.

We found that the resistance value of the PRT sensor changed after gluing it inside the copper sphere. Thus, the PRT sensor installed in the copper sphere was altogether calibrated by the temperature standard group at KRISS with an uncertainty of 12.5 mK. Repeatability and self-heating of the PRT in vacuum were measured to be 2 mK and 5.5 mK, respectively.

Due to the conduction loss through the Manganin wires, the temperature, $T_{\text{BSTP}}$, measured by the BSTP can be different from $T_{\text{atom}}$. Thus, we investigated the dependence of the temperature, especially the conduction loss, on the size and the emissivity of the BSTP by the numerical simulation under simplified thermal environments, as seen in figure 3(a). The outer and inner concentric shells corresponded to the main vacuum chamber and the BBR shield, respectively, and their radii were 100 mm and 50 mm, respectively. The temperature $T_{\text{VC}}$ on the outer spherical shell was fixed at 20 °C, and the temperature $T_{BBRC}$ on the inner shell is at 30 °C. A Manganin wire connected the copper sphere with the outer shell without touching the inner shell through the 4-mm-diameter hole. There are two 10-mm-diameter apertures along the y-axis of the inner shell as in the BBR shield.

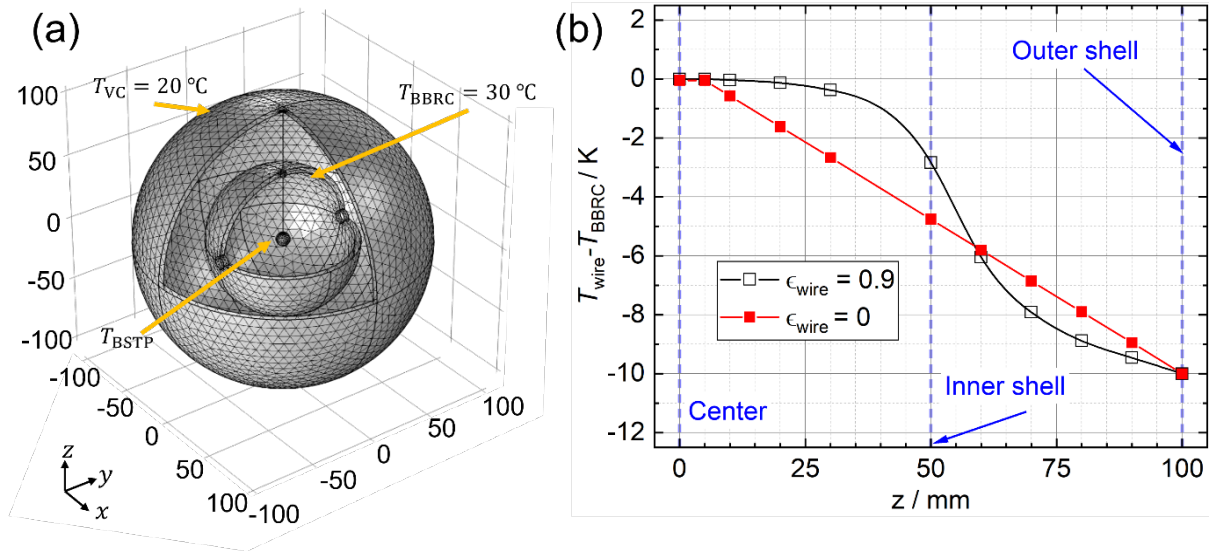

**Figure 3.** (a) Simplified geometry for the numerical simulation to investigate the effects of the geometry and the emissivity of the BSTP on the conductive heat loss. The axis origin is at the center of the spheres. The unit of length is mm. (b) The temperature variation along the wire $T_{\text{wire}}(z)$ was calculated with two different emissivity values of the wire ($\epsilon_{\text{wire}} = 0.9$ for open squares and $\epsilon_{\text{wire}} = 0$ for filled squares) to study the effect of the surface heat radiation on the wire. The BSTP was located at 0 on the z-axis.

Figure 3(b) shows the calculated temperature $T_{\text{wire}}(z)$ along the wire. When we neglected the radiation from the wire surface by setting the emissivity zero, the temperature dropped linearly from the BSTP to the end of the wire at the outer sphere. But with the finite emissivity, the part inside the inner shield absorbed the heat radiation, and the temperature changed more slowly. This radiative absorption significantly reduced the conduction loss from the BSTP, which can be clearly seen in figures 4(a) and 4(b). Figures 4(a) and 4(b) shows the dependence of $T_{BSTP}$ against $T_{BBRC}$ on the emissivity and the radius of the BSTP, respectively, with different values of $\epsilon_{wire}$. $T_{BSTP}$ without the wire corresponds to $T_{atom}$ (represented by black-filled squares in figures 4(a) and 4(b)). As shown in the insets, $T_{atom}$ differed from $T_{BBRC}$ due to the external radiation coming from the outer shell even without the conduction loss.

For a given value of $\epsilon_{wire}$, $T_{BSTP}$ deviated more from $T_{atom}$ for smaller values of the emissivity $\epsilon_{BSTP}$ or radius $r_{BSTP}$ of BSTP. As shown in figure 4(a), with $r_{BSTP} = 5$ mm and $\epsilon_{wire} = 0.9$, the difference of $T_{BSTP}$ from $T_{atom}$ was smaller than 5 mK for $\epsilon_{BSTP} > 0.8$. Also, for given $\epsilon_{BSTP} = 0.9$ and $\epsilon_{wire} = 0.9$, the difference of $T_{BSTP}$ from $T_{atom}$ is smaller than 5 mK for $r_{BSTP} > 4$ mm, as shown in inset of figures 4(b).

Although the larger values of $r_{BSTP}$, $\epsilon_{BSTP}$ and $\epsilon_{wire}$ are of help to make $T_{BSTP}$ be closer to $T_{atom}$, it should be noted that too big $r_{BSTP}$ will not well represent atoms at the trap site. Considering these aspects, we chose the BSTP with $\epsilon_{BSTP} > 0.9$ and $r_{BSTP} = 5$ mm in our experiment. In this condition, the offset of $T_{BSTP}$ from $T_{atom}$ which canbe thought as the black line ,due to the conduction loss by wire was expected to be much smaller than the calibration accuracy of the BSTP, 12.5 mK.

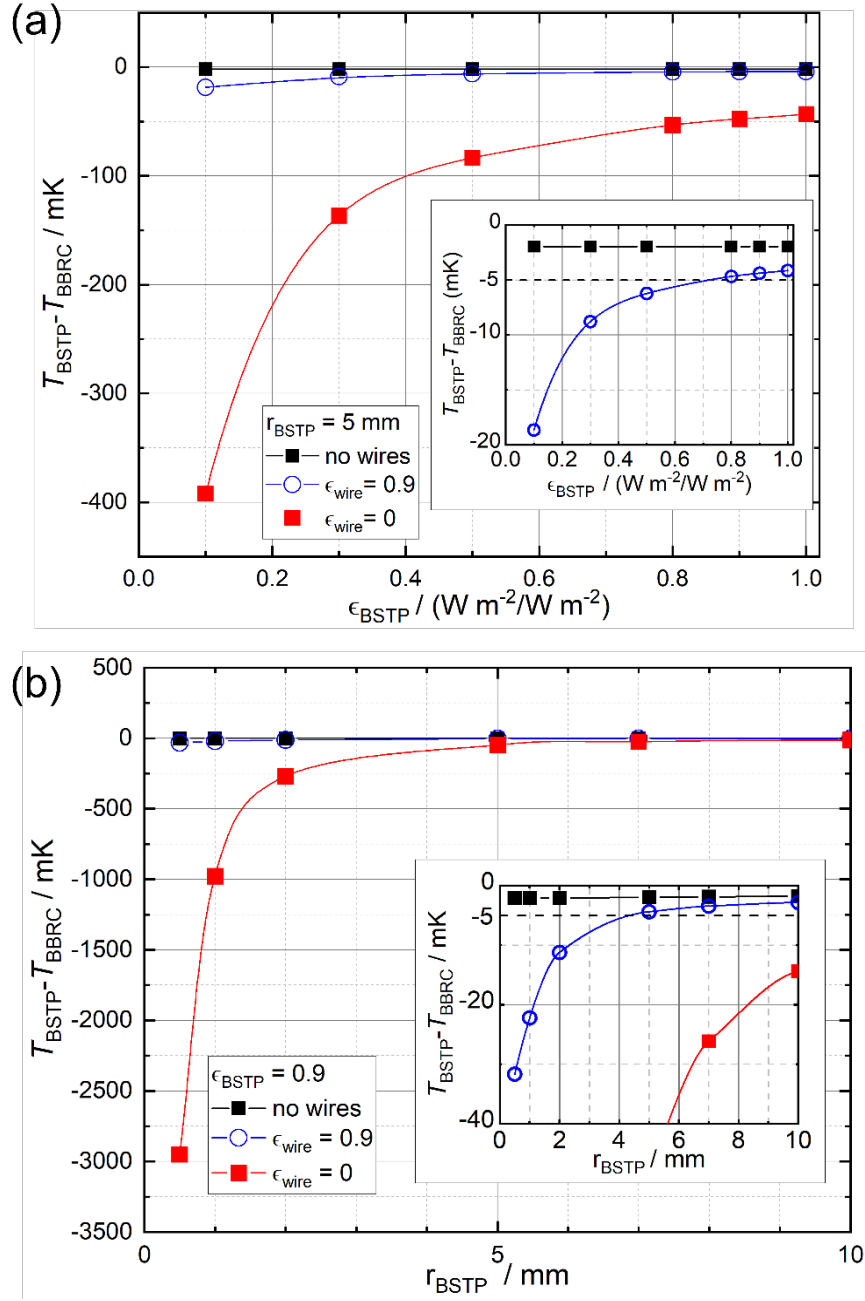

**Figure 4.** (a) The effect of the emissivity $\epsilon_{BSTP}$ of the BSTP and the wire emissivity $\epsilon_{wire}$ on $T_{BSTP}$ with the radius $r_{BSTP}$ of the BSTP fixed at 5 mm. (b) The effect of $r_{BSTP}$ and $\epsilon_{wire}$ on $T_{BSTP}$ with $\epsilon_{BSTP} = 0.9$. Insets are detailed plots near 0 mK. All the curves on data points in the plots are drawn as a guide to the eye.

Next, we experimentally verified the effect of the conduction loss of the BSTP. We changed the temperature $T_{end}$ at the end of the wire by varying the temperature of the electrical feedthroughs connected with the 4-point probe wires of the BSTP using a heater with a PRT from 22 °C to 34 °C as in figure 5. The temperature of the BBR shield was kept at 24 °C, and that of the outer vacuum chamber was around 21.3 °C. The measured sensitivity of $T_{BSTP}$ on the temperature of electrical feedthroughs at the end of 4-point probe wires, was -0.00014(36) K/K as a result of a linear curve fit shown in figure 5. Temperature offset $\Delta T_{conduction}$ of $T_{BSTP}$ from $T_{atom}$ due to conduction loss is

$$\Delta T_{conduction} = -0.00014(36) \times \Delta T_{wire} \qquad (2)$$

, where $\Delta T_{wire} = T_{BSTP} - T_{end} \approx T_{BBR} - T_{end}$.

When $\Delta T_{wire} \approx 10$ K, $\Delta T_{conduction}$ was around 1.4 mK. This is in good agreement with the numerical calculation in figure 4, where $\Delta T_{conduction} \approx 2$ mK for $\epsilon_{wire} = 0.9$, $\epsilon_{BSTP} = 0.9$, $r_{BSTP} = 5$ mm, and $\Delta T_{wire} \approx 10$ K. In the clock operation condition, $\Delta T_{wire} \approx 1$ K, we expect $\Delta T_{conduction} \approx 0.14\,(36)$ mK. The high emissivity of the enamel coating on the wire, $\epsilon_{wire} \approx 0.9$, plays a significant role in reducing the BSTP's conductive loss, and we could neglect the conduction loss of the BSTP for the estimation of $T_{atom}$ from $T_{BSTP}$.

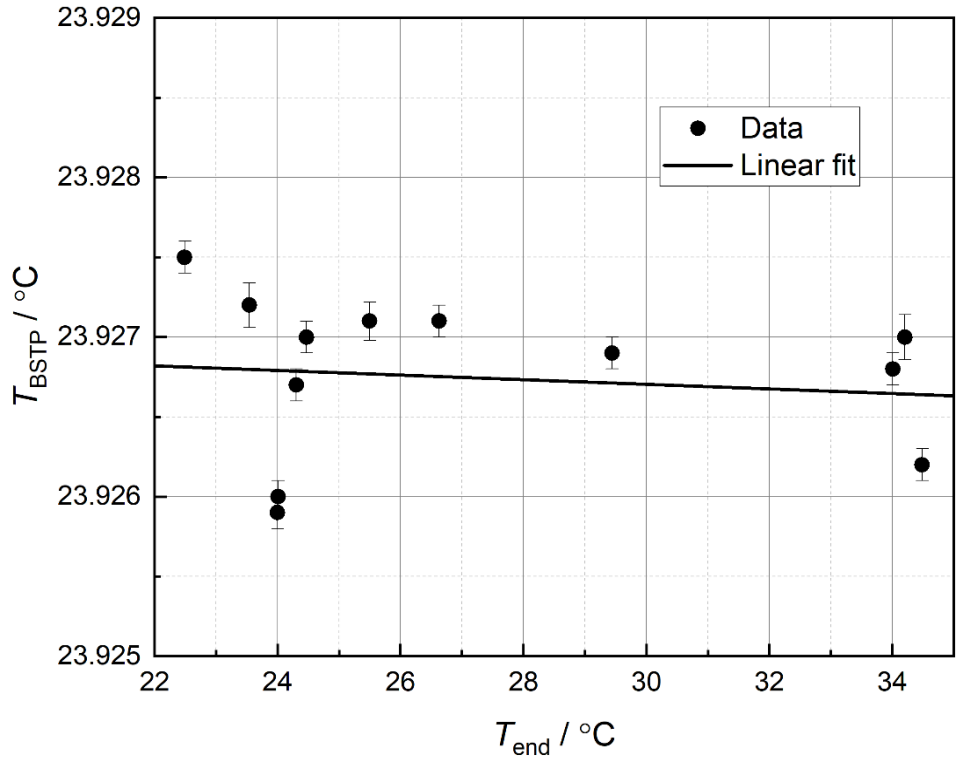

**Figure 5.** The temperature of the BSTP with varying the temperature at the end of the conduction wire. The temperature of the BBR shield was stabilized at 24 °C.

## 4. Measurements of thermal radiations and uncertainty evaluation

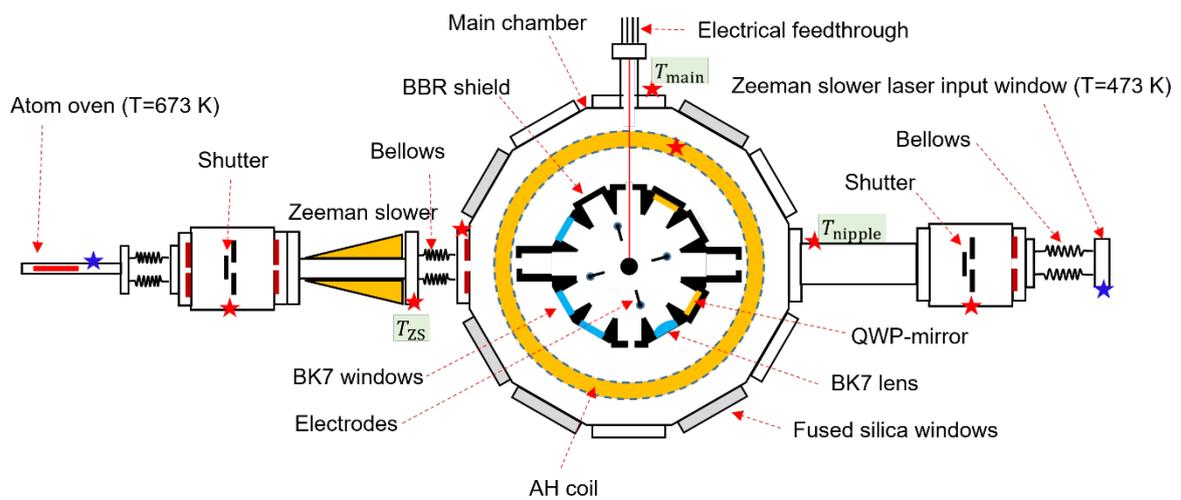

**Figure 6.** Schematic diagram of the whole vacuum system of the Yb optical lattice clock at KRISS. Calibrated PRTs were installed in the positions depicted by red stars and K-type

thermocouple sensors by blue stars. Another PRT, not displayed here, was installed on the front of the main chamber. The heat conductions from the Zeeman slower (ZS) and the Zeeman slower laser input window (ZSLW) were reduced using bellows. The thermal radiations from the atom oven and ZSLW were blocked using shutters. AH coil : Anti-Helmholtz coil.

The whole experimental setup is shown in figure 6 with the BSTP installed. The BSTP will be removed in the actual clock operation. Atoms can see external BBRs from various parts. In general, the temperature at atoms, $T_{\text{atom}}$, can be derived with

$$T_{\text{atom}}^4 = \sum_i \Omega_{eff}^i T_i^4. \tag{3}$$

$\Omega_{\text{eff}}^i$ is a normalized effective solid angle of a heat source $i$, satisfying $\sum_i \Omega_{eff}^i = 1$ and $T_i$ is its temperature. $\Omega_{\text{eff}}^i$ includes the effects of the geometric view factor and the emissivities of all the relevant parts. Dominant heat radiation (and therefore the largest $\Omega_{\text{eff}}^i$) were from the temperature-controlled BBR shield chamber whose temperature was set slightly higher than the main chamber by 1~3 K. In our experimental condition, temperature of each part of the main chamber was similar within 0.5 K and could be represented by the temperature of the main chamber, $T_{\text{main}}$, which is the average value of temperatures on the top, at the side, and at the front of the main chamber. The radiation from the main chamber has effects on BSTP in several ways. One is the direct radiations through open apertures of the BBR shield and transmitted radiation through BK7 optics on the BBR shield, which is significantly suppressed by BK7 glass. The other indirect effect is absorption and emission by BK7 optics on the BBR shield, which causes temperatures of BK7 optics lower than that of the BBR shield due to poor thermal conductivity of BK7 material and imperfect thermal contact with the BBR shield. All of these contribute to the effective solid angle of $T_{\text{main}}$. In addition, we considered the temperature $T_{\text{nipple}}$ of the nipple at the right of the main chamber separately because its radiation can be transferred directly through the open aperture of the BBR shield. It is noted that the conduction and radiation from the heated atom oven (T = 673 K) were suppressed with the bellow and the shutter [21] which is closed during the clock spectroscopy.

Because $T_{\text{atom}}$ can be estimated from the temperature of the BSTP using equation (3), we have

$$T_{\text{atom}}^4 = (T_{\text{BSTP}} - \Delta T_{\text{self}} - \Delta T_{\text{conduction}})^4 = \sum_i \Omega_{eff}^i T_i^4. \tag{4}$$

$\Delta T_{\text{self}}$ is the self-heating offset, 5.5(2.0) mK, as described previously. To obtain $\Omega_{eff}^i$ we varied experimental conditions (stabilized temperature of the BBR shield chamber, Anti-Helmholtz coil current, Zeeman slower duty cycle, cooling water temperature, etc.) as described in Table 1, and measured $T_i$ and $T_{\text{BSTP}}$ simultaneously and obtained 29 samples of distributed temperature (open circles in figure 7). From the multi-variable nonlinear fit of these samples using equation (4), we obtained $\Omega_{eff}^i$ as in Table 1. Normalized solid angles are given in Table 1 and summed up close to 1, confirming that we did not omit significant heat sources.

Table 1. Experimental conditions and determined normalized solid angles. Numbers not parenthesized in the "Temperature range" column were ranges for data set to obtain solid angles, and those parenthesized were for verification data set.

| Heat sources | Temperature range (°C) | Normalized effective solid angle $\left(\Omega_{eff}^i\right)$ |
|---|---|---|
| BBR shield ($T_{\text{BBR}}$)[a] | 22.0-24.0 (22.0-24.0) | 0.97814(15) |
|  | 20.9-21.8 | 0.00229(85) |

| | (21.3-21.9) | |
| --- | --- | --- |
| Main chamber $(T_{\text{main}})$[c] | 21.3-21.9 (21.2-22.7) | 0.01911(78) |
| Zeeman slower $(T_{\text{ZS}})$[d] | 21.2-33.9 (21.5-34.2) | 0.000116(26) |
| Zeeman slower laser input window $(T_{\text{ZSLW}})$ | 20-200 (20-200) | 0.000041(1) |
| Atom oven | 20-400 (room temperature) | suppressed |
| Total | | 0.9997(12) |
| $T_{BSTP}$ | 21.9-23.9 (22.0-24.0) | |

[a] Varied by the control of the stabilized temperature of the BBR shield.
[b] Varied by anti-helmholtz coil current, cooling water temperature, and room temperature.
[c] Varied by anti-helmholtz coil current, cooling water temperature, and room temperature.
[d] Varied by Zeeman slower coil current and room temperature.

Within the range given in Table 1, the temperature at atoms can be estimated using the following equation,

$$T_{\text{atom}} = \sqrt[4]{\Omega_{eff}^{\text{BBR}} T_{\text{BBR}}^4 + \Omega_{eff}^{\text{nipple}} T_{\text{nipple}}^4 + \Omega_{eff}^{\text{main}} T_{\text{main}}^4 + \Omega_{eff}^{\text{ZS}} T_{\text{ZS}}^4 + \Omega_{eff}^{\text{ZSLW}} T_{\text{ZSLW}}^4} \quad (5)$$

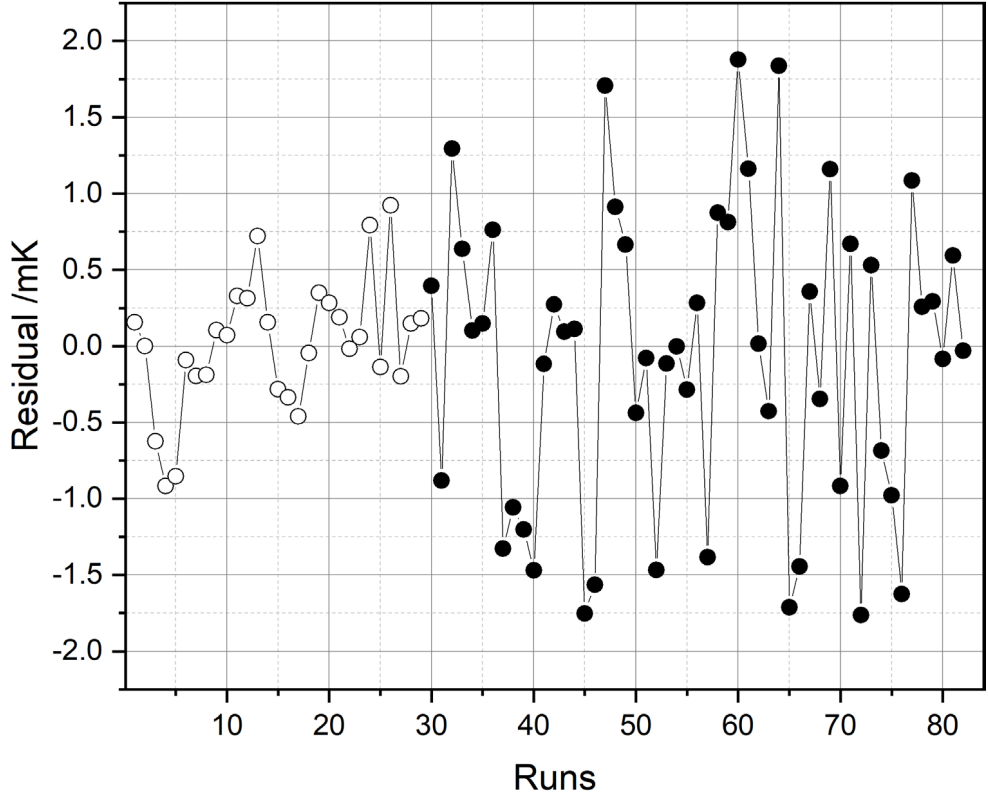

**Figure 7.** The difference between the estimated $T_{\text{atom}}$ from the equation (5) and $T_{\text{atom}}$ estimated from $T_{\text{BSTP}}$ for various experimental conditions. Twenty-nine data points depicted as open cicles were used for the calculation of effective solid angles in Table 1. Black circles were obtained from other sets of experimental conditions. We took 2 mK conservatively as an uncertainty in the prediction of $T_{\text{atom}}$.

The difference between $T_{\text{atom}}$ measured from $T_{\text{BSTP}}$ and the predicted values using equation (5) was $(0.0\pm2.0)$ mK as shown in figure 7 within the experimental condition in Table 1.

The total uncertainty of the atomic temperature under normal operating conditions for the Yb optical lattice clock is given in Table 2.

Table 2. Uncertainty of the atomic temperature ($T_{BBR}$ = 296.15 K, $T_{main}$ = 295.15 K)

| Effects | Correction (mK) | Uncertainty (mK) |
|---|---|---|
| $T_{BSTP}$ calibration | 0.0 | 12.5 |
| $T_{BSTP}$ wire loss (mK/K) | 0.0 | 0.36 |
| $T_{BSTP}$ self-heating | 0.0 | 2.0 |
| Model prediction from equation (5) | 0.0 | 2.0 |
| Total | 0.0 | 13 |

## 5. Conclusion

To evaluate the BBR shift of the Yb optical lattice clock (KRISS-Yb2), we adopted an in-vacuum BBR shield and radiation thermometry using a black-sphere thermal probe. We could obtain the effective solid angles of heat sources experimentally, which did not require exact knowledge of emissivities and geometric view factors of heat sources. To do this, it was important to estimate $T_{atom}$ from the thermal probe measurement. We carried out the numerical simulation of the conduction effect, which implied that high-emissivity surface of the lead wires mitigated the conduction effect significantly, and this was in good agreement with the experimental observation. At the normal operating condition for the Yb optical lattice clock, we obtained the uncertainty of 13 mK, which corresponds to an uncertainty of 0.22 mHz in clock frequency, or $4.2 \times 10^{-19}$ in terms of a fractional frequency. When we combine this with the contribution by the dynamic correction from the atomic response $\Delta \nu_{dyn}$ in equation (1) [2], the total BBR shift uncertainty is $9.5 \times 10^{-19}$. This will pave the way for out Yb optical clock to reach the uncertainty of low $10^{-18}$ level.

## Acknowledgments

The authors thank Young Hee Lee, Inseok Yang, and Wukchul Joung for the calibration of the temperature sensors, and Ki-Lyong Jeong for the transmission measurement of the vacuum view-ports. We thank Kwang Min Yu for the calibration of the multimeter of PRTs. This work was supported by Research on Measurement Standards for Redefinition of SI Units funded by Korea Research Institute of Standards and Science (KRISS – 2022 – GP2022-0001).

## References

[1] Poli N, Oates C W, Gill P and Tino G M 2013 Optical atomic clocks *Riv. del Nuovo Cim.* **36** 555–624

[2] Beloy K, Hinkley N, Phillips N B, Sherman J A, Schioppo M, Lehman J, Feldman A, Hanssen L M, Oates C W and Ludlow A D 2014 Atomic clock with $1\times10^{-18}$ room-temperature blackbody stark uncertainty *Phys. Rev. Lett.* **113** 260801

[3] Sherman J A, Lemke N D, Hinkley N, Pizzocaro M, Fox R W, Ludlow A D, and Oates C W 2012 High-Accuracy Measurement of Atomic Polarizability in an Optical Lattice Clock" *Physical Review Letters*, **108**, 153002

[4] Middelmann T, Lisdat C, Falke S, Vellore Winfred J S R, Riehle F and Sterr U 2011 Tackling the blackbody shift in a strontium optical lattice clock *IEEE Trans. Instrum. Meas.* **60** 2550–7

[5] Dörscher S, Huntemann N, Schwarz R, Lange R, Benkler E, Lipphardt B, Sterr U, Peik E and Lisdat C 2021 Optical frequency ratio of a $^{171}$Yb$^+$ single-ion clock and a $^{87}$Sr lattice clock *Metrologia* **58** 015005

[6]   Kim H, Heo M S, Park C Y, Yu D H and Lee W K 2021 Absolute frequency measurement of the $^{171}$Yb optical lattice clock at KRISS using TAI for over a year *Metrologia* **58** 055007

[7]   Pizzocaro M, Bregolin F, Barbieri P, Rauf B, Levi F and Calonico D 2020 Absolute frequency measurement of the $^1S_0$-$^3P_0$ transition of $^{171}$Yb with a link to international atomic time *Metrologia* **57** 035007

[8]   Hobson R, Bowden W, Vianello A, Silva A, Baynham C F A, Margolis H S, Baird P E G, Gill P and Hill I R 2020 A strontium optical lattice clock with $1 \times 10^{-17}$ uncertainty and measurement of its absolute frequency Richard and *Metrologia* **57** 065026

[9]   Xiong D, Zhu Q, Wang J, Zhang A, Tian C, Wang B, He L, Xiong Z and Lyu B 2021 Finite element analysis of blackbody radiation environment for an ytterbium lattice clock operated at room temperature *Metrologia* **58** 035005

[10]  Schwarz R, Dörscher S, Al-Masoudi A, Benkler E, Legero T, Sterr U, Weyers S, Rahm J, Lipphardt B and Lisdat C 2020 Long term measurement of the $^{87}$Sr clock frequency at the limit of primary Cs clocks *Phys. Rev. Res.* **2** 033242

[11]  Ushijima I, Takamoto M, Das M, Ohkubo T and Katori H 2015 Cryogenic optical lattice clocks *Nat Phot.* **9** 185–9

[12]  Nemitz N, Ohkubo T, Takamoto M, Ushijima I, Das M, Ohmae N and Katori H 2016 Frequency ratio of Yb and Sr clocks with $5 \times 10^{-17}$ uncertainty at 150 seconds averaging time *Nat. Photonics* **10** 258–61

[13]  Ohmae N, Bregolin F, Nemitz N and Katori H 2020 Direct measurement of the frequency ratio for Hg and Yb optical lattice clocks and closure of the Hg/Yb/Sr loop *Opt. Express* **28** 15112–21

[14] Beloy K, Zhang X, McGrew W F, Hinkley N, Yoon T H, Nicolodi D, Fasano R J, Schäffer S A, Brown R C, and Ludlow A D 2018 Faraday-Shielded dc Stark-Shift-Free Optical Lattice Clock *Physical Review Letters*, **120**, 183201

[15] McGrew W F, Zhang X, Fasano R J, Schäffer S A, Beloy K, Nicolodi D, Brown R C, Hinkley N, Milani G, Schioppo M, Yoon T H and Ludlow A D 2018 Atomic clock performance enabling geodesy below the centimetre level *Nature* **564** 87–90

[16] Nicholson T L, Campbell S L, Hutson R B, Marti G E, Bloom B J, McNally R L, Zhang W, Barrett M D, Safronova M S, Strouse G F, Tew W L and Ye J 2015 Systematic evaluation of an atomic clock at $2 \times 10^{-18}$ total uncertainty *Nat. Commun.* **6** 6896

[17] Bothwell T, Kedar D, Oelker E, Robinson J M, Bromley S L, Tew W, Ye J and Kennedy C J 2019 JILA SrI optical lattice clock with uncertainty of $2.0 \times 10^{-18}$ *Metrologia* **56** 065004

[18] Doležal M, Balling P, Nisbet-Jones P B R, King S A, Jones J M, Klein H A, Gill P, Lindvall T, Wallin A E, Merimaa M, Tamm C, Sanner C, Huntemann N, Scharnhorst N, Leroux I D, Schmidt P O, Burgermeister T, Mehlstäubler T E and Peik E 2015 Analysis of thermal radiation in ion traps for optical frequency standards *Metrologia*, **52**, 842-856

[19] Zhang P, Cao J, Shu H-L, Yuan J-B, Shang J-J, Cui K-F, Chao S-J, Wang S-M, Liu D-X and Huang X-R 2017 Evaluation of blackbody radiation shift with temperature-associated fractional uncertainty at $10^{-18}$ level for 40 Ca$^+$ ion optical clock *J. Phys. B*, **50**, 015002

[20] https://www.acktar.com/product/ultra-black/

[21] Kim H, Lee W K, Yu D H, Heo M S, Park C Y, Lee S and Lee Y K 2017 Atom shutter using bender piezoactuator *Rev. Sci. Instrum.* **88** 025101